\documentclass[reprint,letterpaper,
superscriptaddress,
showpacs,
preprintnumbers,
bibnotes,
amsfonts,amsmath,amssymb,
aps,
floatfix,
]{revtex4-1}
\usepackage{graphicx}
\usepackage[caption=false]{subfig}
\usepackage{dcolumn}
\usepackage{bm}
\usepackage{float}
\usepackage[colorlinks=true, linkcolor=blue, citecolor=blue, urlcolor=blue]{hyperref}

\begin{document}

\title{Disorder induced Dirac-point physics in epitaxial graphene from temperature-dependent magneto-transport measurements}

\author{J. Huang}
\author{J. A. Alexander-Webber}
\author{A. M. R. Baker}
\affiliation{\mbox{Department of Physics, University of Oxford, Clarendon Laboratory, Parks Road, Oxford OX1 3PU, United Kingdom}}

\author{T. J. B. M. Janssen}
\affiliation{\mbox{National Physical Laboratory, Hampton Road, Teddington TW11 0LW, United Kingdom}}

\author{A. Tzalenchuk}
\affiliation{\mbox{National Physical Laboratory, Hampton Road, Teddington TW11 0LW, United Kingdom}}
\mbox{\affiliation{Department of Physics, Royal Holloway, University of London, Egham TW20 0EX, United Kingdom}}

\author{\\V. Antonov}
\affiliation{\mbox{Department of Physics, Royal Holloway, University of London, Egham TW20 0EX, United Kingdom}}

\author{T. Yager}
\author{S. Lara-Avila}
\author{S. Kubatkin}
\affiliation{\mbox{Department of Microtechnology and Nanoscience, Chalmers University of Technology, S-412 96 G\"{o}teborg, Sweden}}

\author{R. Yakimova}
\affiliation{\mbox{Department of Physics, Chemistry and Biology (IFM), Link\"{o}ping University, S-581 83 Link\"{o}ping, Sweden}}

\author{R. J. Nicholas}
\email{r.nicholas@physics.ox.ac.uk}
\affiliation{\mbox{Department of Physics, University of Oxford, Clarendon Laboratory, Parks Road, Oxford OX1 3PU, United Kingdom}}

\date{\today}

\begin{abstract}
We report a study of disorder effects on epitaxial graphene in the vicinity of the Dirac point by magneto-transport. Hall effect measurements show that the carrier density increases quadratically with temperature, in good agreement with theoretical predictions which take into account intrinsic thermal excitation combined with electron-hole puddles induced by charged impurities. We deduce disorder strengths in the range 10.2 $\sim$ 31.2 meV, depending on the sample treatment. We investigate the scattering mechanisms and estimate the impurity density to be $3.0 \sim 9.1 \times 10^{10}$ cm$^{-2}$ for our samples. An asymmetry in the electron/hole scattering is observed and is consistent with theoretical calculations for graphene on SiC substrates. We also show that the minimum conductivity increases with increasing disorder potential, in good agreement with quantum-mechanical numerical calculations.
\end{abstract}


\maketitle

Many of the exceptional electronic properties of graphene arise from its linear dispersion relation \cite{r29,r31}. However, when the Fermi energy approaches the Dirac point, its properties can be dominated by the effects of disorder, which can be both intrinsic (such as ripples, topological lattice defects) and extrinsic (including cracks/voids, adatoms, charged impurities, etc.), in general varying from sample to sample \cite{r30}. Of particular significance are the effects of disorder potentials on electrical transport properties \cite{r03} due to the lack of screening at very low carrier densities. Microscopically, the fluctuating electrostatic potential breaks up the intrinsically homogeneous charge distribution into electron-hole puddles \cite{r04,r05,r06,r07,r08}. This effect is recognised to mainly originate from unintendedly introduced charged impurities, whose type, spatial distribution and density also depend on the sample environment, device fabrication techniques, and particularly graphene synthesis and treatment processes. 

Recently, epitaxial graphene on SiC (SiC/G) has been reported to have very high quantum Hall breakdown current density \cite{r27} which potentially allows a quantum electrical resistance standard operating at even higher temperatures and lower magnetic fields \cite{r28}. Low and well controlled carrier density is required to achieve high breakdown current in these conditions, and understanding the disorder effects is therefore highly important. To date, there are very few experimental studies of disorder in epitaxial graphene grown on SiC due to the intrinsically high level of doping from the substrate \cite{r32}. In this letter, using extremely low carrier density epitaxial graphene, we describe the role of disorder in governing the temperature dependent magneto-transport.

Our SiC/G samples were epitaxially grown on the Si-terminated face of 4H-SiC. The devices used in this study all have an 8-leg Hall bar geometry of various sizes fabricated using standard electron beam lithography followed by $\textrm{O}_2$ plasma etching and large-area titanium-gold contacting. A non-volatile polymer gating technique was used to control the carrier density in epitaxial graphene by room-temperature UV illumination \cite{r01} or corona discharge \cite{r02}. Both DC and AC magneto-transport measurements were carried out using an Oxford Instruments 21 T superconducting magnet with a variable temperature insert which allows temperature-dependent measurements from 1.4 K up to 300 K.

\begin{figure*}
	\includegraphics{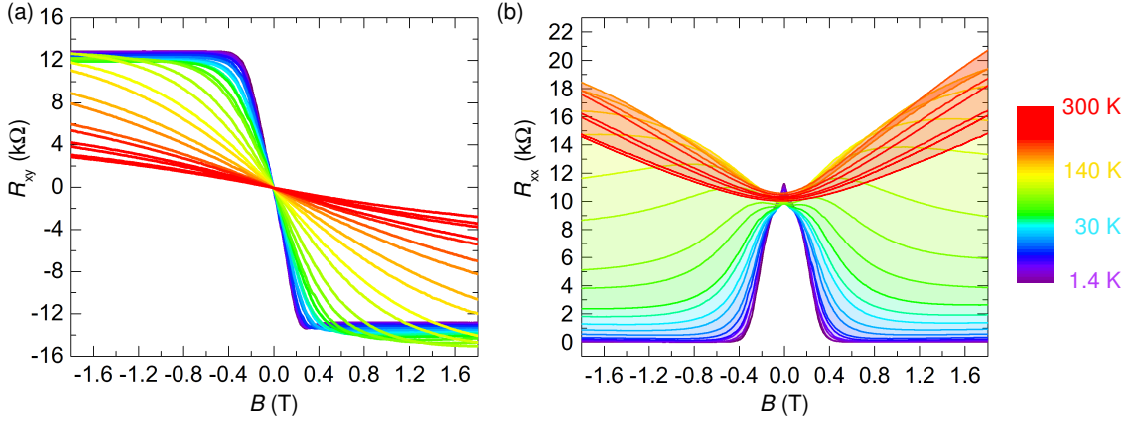}
	\subfloat{\label{FIG1a}}
	\subfloat{\label{FIG1b}}
	\caption{(Color online) The Hall resistance $R_{xy}$ and the longitudinal resistance $R_{xx}$ as a function of magnetic field at temperatures from 1.4 K to 300 K for sample CD2. The sample enters the quantum Hall regime already from about 0.6 T as observed from the quantised $R_{xy}$ and the vanishing $R_{xx}$ at low temperatures.}
	\label{FIG1}
\end{figure*}

Magneto-transport measurements were made on three SiC/G devices, which we denote CD1, CD2 and UV1. We used two different techniques to reduce the relatively high initial electron density and tune the Fermi level to the vicinity of the Dirac point: CD1 and CD2 were treated with multiple negative ion projections produced by corona discharge onto the bilayer polymer gate, resulting in extremely low final electron densities of 1.2 and 1.3 $\times~10^{10}$ cm$^{-2}$, respectively; UV1 was treated with UV illumination which eventually reduced the electron density to 8 $\times~10^{10}$ cm$^{-2}$. As we will show below, these values should not be treated as the real electron densities, but merely are effective carrier densities, $n_{eff}$, calculated from the low-field Hall coefficients at 1.4 K assuming a homogeneous landscape with a single type of charge carriers. Even so, these carrier densities should provide a good, rough estimation of the Fermi energy, which is between 12.7 meV and 32.9 meV, based on the assumption of a linear dispersion where the density of states vanishes at the Dirac point \cite{r26}. In reality, due to the effects of disorder, a residual density of states and coexistence of electrons and holes \cite{r05} at $E_F \rightarrow 0$ are expected, thus, the determination of an extremely low Fermi energy from Hall effect measurements becomes non-trivial. Nevertheless, it is evident that the actual Fermi energy should be lower than the above calculated values. The extremely low effective carrier densities and estimates of the Fermi level show that all measured devices can be treated as being effectively at the Dirac point.

In Fig. \ref{FIG1} we present typical experimental results: the Hall resistance $R_{xy}$ and the longitudinal resistance $R_{xx}$ of sample CD2 as a function of magnetic field at temperatures from 1.4 K to 300 K. In our study, all three devices show similar behaviour as shown in Fig. \ref{FIG1}. Due to the extremely low carrier densities of the samples, quantum Hall plateau corresponding to the filling factor $\nu = 2$ can be observed already from about 0.6 T at 1.4 K. The Hall resistance becomes significantly non-linear when approaching the quantum Hall regime. Therefore, to extract the zero-field carrier densities of our devices, only Hall coefficients between -0.1 T and +0.1 T are used.

It has been theoretically studied and experimentally confirmed that, close to the Dirac point, as a consequence of disorder, the carrier density landscape is extremely inhomogeneous and electron-hole puddles are formed \cite{r03,r04,r05,r06,r07,r08}. Classically, the low-field Hall coefficient in the presence of both electrons and holes is given by,
\begin{equation}
{
R_H \equiv \frac{E_y}{J_x B} = -\frac{1}{e} \frac{n_e \mu_e^2 - n_h \mu_h^2}{(n_e \mu_e + n_h \mu_h)^2}
},
\label{Eq1}
\end{equation}
where $n_e$ ($n_h$) and $\mu_e$ ($\mu_h$) are the electron (hole) density and mobility, respectively. The carrier density directly extracted from the low-field Hall effect is therefore effectively,
\begin{equation}
{
	n_{eff} = \frac{(n_e \mu_e + n_h \mu_h)^2}{n_e \mu_e^2 - n_h \mu_h^2}
}.
\label{Eq2}
\end{equation}
When the Fermi energy is zero, i.e. at charge neutrality point (CNP), $n_e = n_h > 0$. Thus, $n_{eff} = \alpha n_e$, where $\alpha = \frac{\frac{\mu_e}{\mu_h} + 1}{\frac{\mu_e}{\mu_h} - 1}$. Notably, for electron-like behaviour ($R_H < 0$), $\alpha > 0$; for hole-like behaviour ($R_H > 0$), $\alpha < 0$.

\begin{figure}[t]
	\includegraphics{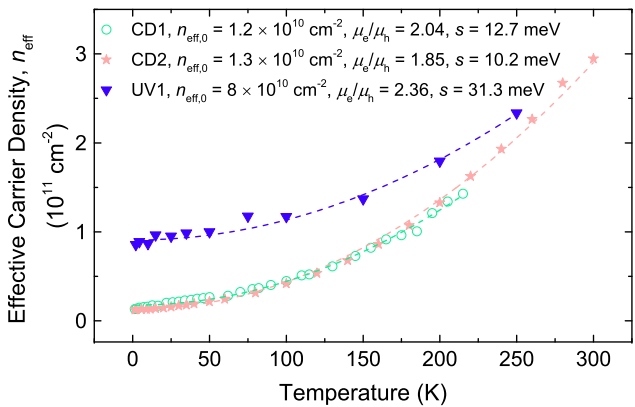}
	\caption{(Color online) Temperature dependence of the effective carrier densities $n_{eff}$ deduced using Eq. (\ref{Eq1}) and (\ref{Eq2}) for sample CD1, CD2 and UV1. Quadratic increase with increasing temperature is observed, together with non-vanishing carrier densities $n_{eff,0}$ at $T \rightarrow$ 0 K. The experimental data is well fitted using Eq. (\ref{Eq4}) and (\ref{Eq5}) as shown in the figure (dash lines), where the disorder potential strength $s$ and the mobility ratio $\mu_e/\mu_h$ are extracted from the fitting.}
	\label{FIG2}
\end{figure}

We now analyse the temperature dependence of the effective carrier density $n_{eff}$ as shown in Fig. \ref{FIG2} for the three devices. A quadratic increase of $n_{eff}$ with increasing temperature can be clearly observed for all of the samples. Each sample also exhibits a distinct non-zero residual charge density at the low temperature limit even when $E_F \rightarrow 0$, indicating that the potential landscape of our devices is highly inhomogeneous. These features are clearly different from the Arrhenius behaviour of conventional semiconductors and intrinsic thermal activation in graphene when no disorder effects are accounted for (i.e., there is no residual carrier density). Accurate fitting can be made based on the theory \cite{r03} assuming that the electronic potential energy of disordered graphene follows Gaussian statistics, which give the probability of finding the local potential within a range $dV$ about $V$,
\begin{equation}
{
	P(V) dV = \frac{1}{\sqrt{2 \pi s^2}} e^{-\frac{V^2}{2s^2}} dV
},
\label{Eq3}
\end{equation}
where $s$ is a parameter used to characterise the strength of the potential fluctuations. As a consequence, the temperature-dependent charge density at CNP for both electrons and holes are \cite{r03},
\begin{equation}
{
	n_e(T) = n_h (T) = \frac{g_sg_v}{2 \pi (\hbar v_F)^2} [\frac{s^2}{4} + \frac{(\pi k_B T)^2}{12}] 
},
\label{Eq4}
\end{equation}
where $g_s = g_v = 2$ are the spin and valley degeneracies, and $v_F \approx 10^6$ m/s is the Fermi velocity. The temperature dependence of the effective carrier density is therefore,
\begin{equation}
{
	n_{eff}(T) = \alpha n_e(T)
},
\label{Eq5}
\end{equation}
where $\alpha$ is assumed to be constant over the temperature range under consideration. The predicted temperature dependence from Eq. (\ref{Eq4}) and (\ref{Eq5}) fits the experimental data very well (Fig. \ref{FIG2}), giving potential fluctuation strengths $s =$ 12.7, 10.2 and 31.3 meV, and pre-factors $\alpha$ which translate into mobility ratios of electrons to holes $\mu_e/\mu_h =$ 2.04, 1.85 and 2.36, for the devices CD1, CD2 and UV1 respectively. Table \ref{Tab1} shows comparisons of the potential fluctuations due to electron-hole puddles, between the values deduced from our magneto-transport measurements and those found in the literature \cite{r04,r05,r06,r07,r08}, where most of the characterizations are based on scanning tunnelling microscopy (STM). Table \ref{Tab1} also includes the disorder strength (15 $\pm$ 1 meV) from our analysis of the published data for SiC/G samples exposed to aqueous-ozone (AO) processing \cite{r41}, which results in high mobility and extremely low p-type doping with an effective carrier density $n_{eff,0} = - 4.0 \times 10^{10}$ cm$^{-2}$ (negative sign for hole-like behaviour) from Hall measurements. We find that the disorder strengths measured in our samples are consistent with those reported previously for SiC/G, and are smaller than those of CVD and exfoliated samples on SiO$_2$, while they are slightly larger than that of exfoliated graphene on h-BN, which is an atomically smooth, dangling bonds free and lattice-matched substrate to support high quality graphene \cite{r09}. These comparisons suggest that SiC/G generally has very good quality and relatively small amounts of disorder, even though the actual characteristics are expected to vary from sample to sample and may also be sensitive to the sample treatment as seen from Table \ref{Tab1}. At the same time, it is demonstrated that magneto-transport measurement is an additional effective method to investigate the disorder effects and characteristics in graphene.

\begingroup
\squeezetable
\begin{table}[t]
	\caption{Energy Fluctuations of e-h Puddles in Graphene}
	\begin{ruledtabular}
		\begin{tabular}{lcc}
			\noalign{\smallskip}
			\textrm{Synthesis (Treatment)}&
			\textrm{Disorder Strength}&
			\textrm{Probing Method}\\
			\noalign{\smallskip}
			\hline
			\noalign{\smallskip}
			\noalign{\smallskip}
			Epitaxial on SiC (\textbf{CD1}) & \textbf{12.7 $\pm$ 0.6 meV} & \textbf{Magneto-transport}\\
			\noalign{\smallskip}
			Epitaxial on SiC (\textbf{CD2}) & \textbf{10.2 $\pm$ 0.4 meV} & \textbf{Magneto-transport}\\
			\noalign{\smallskip}
			Epitaxial on SiC (\textbf{UV1}) & \textbf{31.3 $\pm$ 2.0 meV} & \textbf{Magneto-transport}\\
			\noalign{\smallskip}
			Epitaxial on SiC (\textbf{AO}) & \textbf{15 $\pm$ 1 meV} & \textbf{Magneto-transport}\\
			\noalign{\smallskip}
			Epitaxial on SiC & 12 meV & KPM \cite{r04}\\
			\noalign{\smallskip}
			Exfoliated on SiO$_2$/Si & 50 meV & SET \cite{r05}\\
			\noalign{\smallskip}
			Exfoliated on SiO$_2$/Si & $\sim$~20 meV & STM \cite{r06}\\
			\noalign{\smallskip}
			Exfoliated on h-BN & 5.4 meV & STM \cite{r07}\\
			\noalign{\smallskip}
			CVD on Ir(111) & $\sim$~30 meV & STM/STS \cite{r08}\\
			\noalign{\smallskip}
		\end{tabular}
	\end{ruledtabular}
	\label{Tab1}
\end{table}
\endgroup

\begin{figure}
	\includegraphics{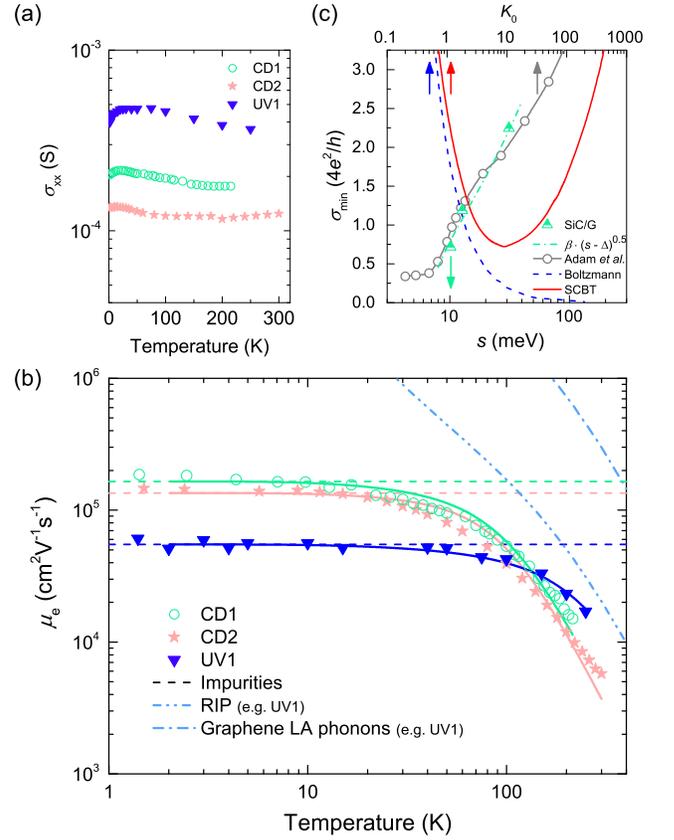}
	\subfloat{\label{FIG3a}}
	\subfloat{\label{FIG3b}}
	\subfloat{\label{FIG3c}}
	\caption{(Color online) (a) The longitudinal conductivity as a function of temperature, where weak non-monotonic dependences are shown. (b) The temperature dependence of the electron mobility of our samples. Individual contributions due to impurity scattering (green/pink/blue dash lines) for all three samples, LA phonon scattering (blue dash-dot line), RIP scattering (blue dash-dot-dot line) for UV1 as an example are shown. The solid lines represent the overall $\mu_e(T)$ dependence by fitting the experimental data. (c) $\sigma_{min}$ as a function of disorder strength. An $\beta (s-\Delta)^{\frac{1}{2}}$ dependence (green dash-dot line) is observed from our experimental data (green triangles). $\sigma_{min}$ as a function of $K_0$ from numerical calculations by Adam \textit{et al.} (gray circles and solid line), as well as predictions using the Boltzmann (blue dash line) and the self-consistent Boltzmann (red solid line) theories are also shown \cite{r25}.}
	\label{FIG3}
\end{figure}

To evaluate the scattering mechanisms in our SiC/G samples, we now turn to examine the temperature dependence of the longitudinal conductivity $\sigma_{xx}$ and the electron mobility as shown in Fig. \ref{FIG3}. Carrier mobilities of individual species are calculated classically based on,
\begin{equation}
{
	\sigma_{xx} = e (n_e \mu_e + n_h \mu_h) = \frac{e n_{eff}}{\alpha}(\mu_e + \mu_h)
},
\label{Eq6}
\end{equation}
via Eq. (\ref{Eq5}) and the value $\frac{\mu_e}{\mu_h}$ deduced from $\alpha$. It is observed that $\sigma_{xx}(T)$ remains slowly varying with weak non-monotonic fluctuations for a large range of temperatures. Similar behaviour has been reported for monolayer graphene samples when $E_F \approx 0$ \cite{r10,r11}, and this is clearly different from thermally activated conductivity in conventional gapped semiconductors and from phonon-limited behaviour in graphene, which will result in a $T^{-4}$ or $T^{-1}$ dependence \cite{r14,r15} at low or high temperatures respectively due to intravalley acoustic phonon scattering. It should be pointed out that this temperature dependence of conductivity in our extremely low carrier density samples could be a combination of various contributions. It is believed that this weakly varying conductivity is mainly due to the temperature-dependent carrier density as described above and the $\mu(T)$ dependence as we will discuss below. At the lowest temperatures there are also temperature dependent weak localization corrections, which can be seen from Fig. \ref{FIG1b} around $B = 0$ T but have been excluded in Fig. \ref{FIG3a}. Fig. \ref{FIG3b} shows the electron mobility as a function of temperature, as well as the mobility limits as a result of various scattering mechanisms, including impurity scattering, scattering by longitudinal acoustic (LA) phonons in graphene, and also by remote interfacial phonons (RIP) at the SiC/graphene interface \cite{r12,r13}. In the case of charged impurities, carrier mobility is inversely proportional to the impurity density $n_{imp}$ \cite{r21,r26},
\begin{equation}
{
		\mu_{imp} \approx \frac{C_0}{n_{imp}}
},
\label{Eq7}
\end{equation}
where $C_0$ is a constant.
For LA phonon scattering \cite{r14},
\begin{equation}
{
	\mu_{LA} = \frac{e \hbar \rho_s v_s^2 v_F^2}{\pi n_e D_A^2 k_B T}
},
\label{Eq8}
\end{equation}	
with $\rho_s =$ 7.6 $\times~10^{−7}$ kg/m$^2$ the two-dimensional mass density, $v_s =$ 1.7 $\times~10^{4}$ m/s the sound velocity, and $D_A =$ 18 eV the acoustic deformation potential. The RIP limited mobility is given by \cite{r12,r13}, 
\begin{equation}
{
	\mu_{RIP} = \frac{1}{n_e e} [\sum\limits_{i}^{} (\frac{C_i}{\exp{(\frac{E_i}{k_B T})} - 1})]^{-1}
},
\label{Eq9}
\end{equation}
where $C_i$ and $E_i$ are electron-phonon coupling constants and phonon energies of the phonon modes under consideration. To fit our data, we first considered three phonon modes: two out-of-plane acoustic phonon modes in epitaxial graphene ($E_1 = 70$ meV and $E_2 = 16$ meV) \cite{r12, r37} and a surface phonon mode of 4H-SiC ($E_3 = 117$ meV) \cite{r12,r13,r38}. However, due to their relative large phonon energies, none of these can yield a reasonable fit, which can only be obtained (Fig. \ref{FIG3b} solid lines) when an additional low-energy phonon mode ($E_4 \approx 2$ meV) is introduced. This is consistent with the previously reported results \cite{r12,r13,r39,r40}, and this low-frequency remote phonon mode has been recognised to originate from the interaction between graphene and the buffer layer, that they are oscillating out-of-phase parallel to each other.

It can be seen from Fig. \ref{FIG3b} that impurity scattering plays the most dominant role at low temperatures ($<$ 100 K), while the high-temperature mobility is probably limited by RIP scattering, since LA phonons make only a small contribution to the overall mobility for temperatures below 400 K. Using $C_0 \approx 5 \times 10^{15}$ V$^{-1}$s$^{-1}$ \cite{r21}, the densities of charged impurities for our SiC/G samples are estimated to be $3.0 \sim 9.1 \times 10^{10}$ cm$^{-2}$, which are 1 $\sim$ 2 orders of magnitude lower than that in typical exfoliated \cite{r33} and CVD grown \cite{r34,r35} graphene on SiO$_2$, but are comparable to that of h-BN supported graphene \cite{r36}, consistent with its high charge carrier mobility. Even though we restrict the above analysis to phonon and impurity scattering, other possible scattering mechanisms exist, such as scattering due to ripples \cite{r18,r19} and very large defects \cite{r17}. Quantitative analysis of these mechanisms on our devices is rather difficult since systematic examination of the sample morphology is required and, on the other hand, the theoretical pictures are rather complicated and still contentious \cite{r16}.

So far we have been able to identify that charge carrier scattering at low temperatures in our SiC/G is mainly due to impurities, in the classical regime where quantum corrections are suppressed by magnetic fields. It is these impurities which provide the same origin for generating the electron-hole puddles at $E_F \rightarrow 0$. Furthermore, we believe that these impurities are most likely to be charged/Coulomb impurities rather than short-range impurities. The main evidence for this is the presence of unequal electron and hole mobilities, which is a consequence of the unbalanced transport cross sections due to charged scatterers in a system with 2D relativistic dispersion \cite{r20,r21}. This can be intuitively understood as because an attractive potential scatters a charge carrier more effectively than a repulsive potential \cite{r20}. As presented in Fig. \ref{FIG2}, we have obtained similar $\mu_e/\mu_h$ in the range of 1.85 $\sim$ 2.36. According to the theory \cite{r20}, assuming a single species of monovalent ($|Z| = 1$) impurities, the above mobility ratios can be translated into a dimensionless asymmetry factor $c = 0.30 \sim 0.39$, which is used to characterise the strength of this asymmetry effect (i.e. $c = 0$ for $\mu_e = \mu_h$ and $c \rightarrow 1$ for $\mu_{e(h)} \gg \mu_{h(e)}$). The nature of this asymmetry factor depends on the dielectric constant of the substrate: for SiO$_2$, $c|_{\epsilon_r = 3.9} \approx 0.46$; for SiC substrates, the same as used in our devices, $c|_{\epsilon_r = 10.0} \approx 0.32$, which is in very good agreement with our experimental results. Small variations around the predicted value are expected, since the actual electrostatic environment of each SiC/G sample could also be affected by the polymer top-gate dielectrics, meanwhile, the types and amounts of charged impurities present in our samples could be more complex.

Finally, the effects of disorder potential fluctuations on the low-temperature non-vanishing minimum conductivity ($\sigma_{min}$) at the Dirac point are investigated for graphene in the diffusive transport regime. This property has been extensively considered theoretically and the two main existing approaches lead to contradictory results \cite{r25}. The semiclassical Boltzmann transport theory predicts a decreasing $\sigma_{min}$ with increasing disorder strength. With a self-consistent modification to the Boltzmann theory, a subsequent increase of the minimum conductivity for higher disorder strengths is predicted. On the other hand, the minimum conductivity treated quantum-mechanically \cite{r22,r23,r24,r25} is increased for the entire disorder strength range for a non-interacting model using a Gaussian correlated disorder potential. Experimentally, very few studies can be found addressing this problem in the literature \cite{r21}. Shown in Fig. \ref{FIG3c} is the minimum conductivity (at $B = 0$) as a function of the disorder strength $s$ obtained from our measurements when quantum corrections have been taken into account, as well as theoretical predictions including the numerical calculation via the quantum mechanical approach by Adam \textit{et al.} \cite{r25}, and results form the (self-consistent) Boltzmann theories, for $L = 50 \xi$, where $L$ is the sample length, $\xi$ is the correlation length of the assumed random Gaussian potential $U(\textit{\textbf{r}})$ in the system and the dimensionless parameter $K_0 \propto \langle U(\textit{\textbf{r}})U(\textit{\textbf{r}}') \rangle$ is the disorder strength used in the theories. Our experimental results show that the minimum conductivity increases with increasing $s$, roughly following a $\beta (s-\Delta)^{\frac{1}{2}}$ dependence locally in the (0.5 $\sim$ 2.5) $\times~\frac{4 e^2}{h}$ range, highlighted by the green dash-dot line in the figure, where $\beta$ and $\Delta$ are constants. This increase agrees qualitatively well with the theoretical predictions \cite{r25} from the quantum-mechanical approach, where we assume $s \propto \sqrt{K_0}$. However, our data do not agree with the results from the Boltzmann and the self-consistent Boltzmann theory as shown in the figure. In addition, we note that the minimum conductivity may have a complex dependence on the sample length and details of quantum interference effects \cite{r25,r42,r43}, and also be a function of the charged impurity density $n_{imp}$ indicated from previous experimental work by Chen \textit{et al.} \cite{r21}, whose results suggest that $\sigma_{min}$ drops with increasing $n_{imp}$ at low impurity densities and may saturate rapidly. To allow a more conclusive interpretation, however, more experimental data and systematic comparisons between well-controlled samples from different synthesis methods and a larger range of disorder potentials and impurity densities would be needed.

In summary, we have presented temperature dependent magneto-transport measurements on epitaxial graphene. We have demonstrated the disorder effects when the Fermi energy lies in the vicinity of the Dirac point, and have been able to identify the main origin of those effects to be charged impurities. The disorder strength and the impurity densities of our samples have been estimated from experimental results. We have also shown that the minimum conductivity increases with increasing disorder strength, in good agreement with numerical quantum-mechanical calculations. Overall, the application of this method can, therefore, provide an alternative and effective route for quantitatively studying the disorder characteristics in graphene and other two-dimensional materials.

This work was partially supported by the U.K. EPSRC and NMS, E.U. Graphene Flagship (Contract No. CNECT-ICT-604391), and EMRP GraphOhm. 
Work at Chalmers University of Technology was supported by the Swedish Foundation for Strategic Research (SSF), Linnaeus Centre for Quantum Engineering, Knut and Alice Wallenberg Foundation and Chalmers AoA Nano.
J. H. acknowledges financial support from the China Scholarship Council.

\nocite{*}
\bibliography{refs}

\end{document}